\documentclass[twocolumn,a4paper, traditabstract]{aa}

\usepackage{amsmath}
\usepackage{graphicx,txfonts,natbib,verbatim}
\usepackage{mathrsfs,amsfonts,amssymb}
\usepackage{color, ulem}
\bibpunct{(}{)}{;}{a}{}{,}

\begin{document}

\title{Large scale simulations of solar type III radio bursts: flux density, drift rate, duration and bandwidth}
\author{H.~Ratcliffe\inst{\ref{inst1}, \ref{inst2}}\and E.P.~Kontar\inst{\ref{inst2}}\and H.A.S.~Reid\inst{\ref{inst2}} }

\institute{Centre for Fusion, Space and Astrophysics, Department of Physics, University of Warwick, CV4 7AL, United Kingdom
\label{inst1} \and SUPA School of Physics \& Astronomy, University of Glasgow, G12 8QQ, United Kingdom \label{inst2}}

\offprints{H. Ratcliffe \email{h.ratcliffe@warwick.ac.uk}}

\date{Received ; Accepted }

\abstract{Non-thermal electrons accelerated in the solar corona can produce intense coherent
radio emission, known as solar type III radio bursts. This intense radio emission is often observed from hundreds of MHz in the corona down to the tens of kHz range in interplanetary space.
It involves a chain of physical processes from the generation of Langmuir waves to nonlinear
processes of wave-wave interaction. We develop a self-consistent model to calculate radio
emission from a non-thermal electron population over large frequency range,
including the effects of electron transport, Langmuir wave-electron interaction, the evolution
of Langmuir waves due to non-linear wave-wave interactions, Langmuir wave conversion
into electromagnetic emission, and finally escape of the electromagnetic waves.
For the first time we simulate escaping radio emission over a broad frequency range from 500~MHz down
to a few MHz and infer key properties of the radio emission observed: the onset (starting) frequency, {identification as fundamental or harmonic emission}, peak flux density, instantaneous frequency bandwidth, and timescales for rise and decay. Comparing with the observations, these large scale simulations enable us to identify
the processes governing the key type III solar radio burst characteristics.}

\keywords{Sun:particle emission, Sun: radio radiation, Sun:flares}
\titlerunning{Solar type III radio burst simulations}
\authorrunning{Ratcliffe {\it et al.}}

\maketitle

\section{Introduction}

Type III solar radio bursts, easily identified due to their high  brightness temperatures
and rapid frequency drift, are a common signature of fast electron beams in the solar
corona. It is known that they occur due to accelerated electrons propagating in the decreasing
density plasma of the corona and solar wind, which generate Langmuir waves
and subsequently electromagnetic emission at the plasma frequency
and its harmonics. The frequency of these bursts {therefore} traces the local plasma frequency, {and so information about the electron beam velocity can be inferred if the density profile is known.} However, due to the complicated nature of particle transport and emission,
it is not straightforward to recover any additional information about the electrons.
Nevertheless, type III observations have the potential to offer unique information
supplementary to that from e.g Hard X-Rays (HXR) \citep[e.g.][as a review]{1988SoPh..118...49D,2011SSRv..159..107H}
and gyrosynchrotron radio emission observations \citep[e.g.][]{1990ApJ...361..290G,2002ApJ...580L.185M,2013ApJ...768..190F},
in order to understand electron acceleration in the corona.

The plasma emission mechanism, which we assume is responsible for  type III burst production, was first proposed by \citet{1958SvA.....2..653G} and
is summarised in the books by \citet{1980MelroseBothVols, 1995lnlp.book.....T},
and reviews by \citet{1985ARA&A..23..169D, 1998ARA&A..36..131B, 2008SoPh..253....3N,2009IAUS..257..305M,2014arXiv1404.6117R}.
Large amounts of work have been invested in the problem yet the exact
details of type III production are still not fully understood. The mechanism involves multiple steps,
several of which are non-linear, and these must all be considered self-consistently.
Broadly, one needs to take into account four elements
leading to escaping radio emission. Firstly, the spatially non-uniform electron beam propagates through the coronal and solar wind plasma and generates Langmuir waves;
secondly, the Langmuir waves evolve both spectrally and in space;
thirdly, electromagnetic waves are generated by the Langmuir waves
and finally, this radio emission escapes the source and reaches the observer.

{These main processes operate on similar scales, requiring a self-consistent treatment, and so} numerical simulations are essential
to reproduce the basic burst properties, such as drift rate, exciter speed, frequency
bandwidth and rise/decay times. Specifically the consideration of electron transport including
the generation and absorption of Langmuir waves
requires a numerical treatment to achieve self-consistency.
Langmuir waves generated by the beam electrons are subsequently reabsorbed by the beam \citep[e.g.][]{1972SoPh...24..444Z,1976SoPh...46..323T,1977SoPh...55..211M,1999SoPh..184..353M},
which allows the electrons to propagate over more than 1~AU accompanied
by a high level of plasma waves.
The presence of plasma inhomogeneity substantially
complicates the evolution of these Langmuir waves \citep[e.g.][]{1979ApJ...233..998S,1982PhFl...25.1062G,1987SoPh..111...89M,2005PhRvL..95u5003Y,2011PhPl...18e2107D} as can be seen in numerical simulations
\citep[e.g.][]{2001A&A...375..629K,2004SoPh..222..299L,2012ApJ...761..176R}.
The overall level of Langmuir waves induced by the beam is decreased but the scattering of Langmuir waves to lower wavenumber can lead to acceleration of electrons.

Large-scale 1-D numerical simulations of electron transport, including the effects
of Langmuir wave generation, have recently been developed
\citep{2009ApJ...695L.140K, 2010ApJ...721..864R,2013SoPh..285..217R}, and show significant effects
on the electron spectra below $\sim 50$~keV observed at 1~AU. Specifically, an injected
power-law spectrum develops into a broken power-law, and the low energy electrons arrive
sooner than their high-energy counterparts, both of which effects are due to Langmuir wave
evolution in the decreasing density plasma of the corona and heliosphere.
Simulations of type III burst production have also recently been developed for rather
narrow frequency range by e.g. \citet{2006PhRvL..96n5005L,2008JGRA..11306104L, 2011ApJ...738L...9L,2013JGRA..118.4748L},
and can reproduce some features of observed type III dynamic spectra.

In this paper we present type III burst simulations based on the description of electron propagation
and Langmuir wave evolution in inhomogeneous plasma of \citet{2013SoPh..285..217R}.
We add an angle-averaged model for plasma radio emission, described in \citet{2014A&A...562A..57R},
in order to simulate the type III burst generation process from electrons to radio waves.
We calculate the duration
and the frequency bandwidth of emission over a wide frequency range. The frequency
drift of the emission is used to infer the velocity of the emission source,
which may then be compared to the true velocity of the generating electrons.
We discuss the physical origins of these major
characteristics, allowing clarification of the details of the standard model of emission
and ruling out some suggestions as to dominant effects. Finally we also consider the characteristics
of the fast-electron injection in the corona, specifically the injection time-scale,
and the effects this has on the observed emission.

\section{Simulation Model}

\subsection{Electron-Langmuir wave interactions}\label{sec:elecLang}

 We start from the simulations of beam propagation and Langmuir wave generation described
in \citet{2013SoPh..285..217R}, which treat the propagation and evolution of a fast electron population streaming
from the Sun outwards through the expanding  corona.
The model assumes that the dynamics of electrons and Langmuir waves is one-dimensional along the direction
of ambient magnetic field. Because the magnetic field expands with distance (spherically symmetrically)
the electron beam has a cross-sectional area increasing with distance.
The whole system has azimuthal symmetry around the direction of ambient magnetic field.
The quasi-linear equations describing the electron and Langmuir wave evolution are based
on those in \citet{drummond1962nucl, vedenov1962quasi}, but modified to account for additional
processes described below:
\begin{align}
\frac{\partial f}{\partial t} + \frac{{\rm v}}{(r+r_0)^2}\frac{\partial}{\partial r}(r + r_0)^2f &=
\frac{4\pi ^2e^2}{m_e^2}\frac{\partial }{\partial {\rm v}}\frac{W}{{\rm v}}\frac{\partial f}{\partial {\rm v}}\notag  \\& +\frac{4\pi n_e e^4}{m_e^2}\ln\Lambda\frac{\partial}{\partial {\rm v}}\frac{f}{{\rm v}^2}+S({\rm v}, r,t),
\label{eqn:k1}
\end{align}
\begin{align}
\frac{\partial W}{\partial t} &+ \frac{\partial \omega_L}{\partial k}\frac{\partial W}{\partial r}
-\frac{\partial \omega _{pe}}{\partial r}\frac{\partial W}{\partial k}
= \frac{\pi \omega_{pe}(r)}{n_e}{\rm v}^2W\frac{\partial f}{\partial {\rm v}} \notag\\& +
e^2\omega_{pe}(r) {\rm |v|} f \ln{\frac{{\rm |v|}}{{\rm v}_{Te}}}
- (\gamma_{c} +\gamma_L )W  + {\rm St}_{decay}(W)+{\rm St}_{ion}(W),
\label{eqn:k2}
\end{align}
where  $f({\rm v},r,t)$ is the electron beam distribution function, and $W(k,r,t)$ the spectral energy density
of Langmuir waves, with $k$ their wavenumber.  $S({\rm v}, r,t)$ is the electron injection
rate discussed in Section \ref{sec:InitCond}.

{The first two terms on the RHS of Equations (\ref{eqn:k1}) and (\ref{eqn:k2}) describe the
interaction of Langmuir waves and electrons. As in \citet{2013SoPh..285..217R}, the electron transport assumes radial
expansion of the beam due to expansion of magnetic field in the corona (described by the second term on the RHS of
Equation (\ref{eqn:k1})). The ambient magnetic field is assumed to expand in a cone, with opening
angle $34^\circ$ determined by the constant $r_0=3.4\times10^9$~cm in the second term
on the  LHS of Equation (\ref{eqn:k1}) \citep{2013SoPh..285..217R}.
The cross-sectional area of the electron beam therefore expands, and the electron density correspondingly
decreases. As the Langmuir wave group velocity is a fraction of the electron thermal speed,
the Langmuir waves are generated/absorbed locally and the role of {magnetic flux} expansion on Langmuir waves
is negligible.  The electrons and Langmuir waves interact resonantly, $\omega_{pe}=k {\rm v}$,
so that the electrons with velocity ${\rm v}$ interact with Langmuir waves of equal phase velocity $\omega_{pe}/k$.
The second term on the RHS of Equation (\ref{eqn:k1}) also accounts for the (total) electron energy losses in plasma.}

{The electron distribution function $f({\rm v},r,t)$ and the spectral energy density of Langmuir waves $W(k,r,t)$ are one-dimensional functions, with positive direction of ${\rm v}$ and $k$ corresponding to the direction away from the Sun. In the following sections we also distinguish ``forwards'' (positive $k$)
and ``backwards'' Langmuir waves, that is those with a component of their wavevector away from the Sun,
and those towards the Sun.}

A small part of the energy lost by electrons to the plasma (second term on the RHS of
Equation (\ref{eqn:k1})) goes to generation of Langmuir waves near $\omega _{pe}$.
This so called spontaneous Langmuir wave generation (the second
term on the RHS of Equation (\ref{eqn:k2})), is treated as in e.g. \citet{1970SvA....14...47Z,1976SoPh...46..323T,2012A&A...539A..43K}.
The third term on the LHS of Equation (\ref{eqn:k2})  describes the change in wavenumber
of Langmuir waves due to density gradients
in the background plasma. The third term on the RHS describes collisional absorption
of Langmuir waves by background plasma \citep{1981phki.book.....L}
with coefficient $\gamma_c
= 4\sqrt{2\pi} n_e e^4\ln\Lambda/(3m_e^2 {\rm v}_{Te}^3) \simeq
\pi n_e e^4\ln\Lambda/(m_e^2 {\rm v}_{Te}^3)$ with the Coulomb logarithm $\ln\Lambda$ taken as 20 for
the solar corona. The background plasma electrons absorb Langmuir waves due to Landau damping in Maxwellian plasma with
coefficient $\gamma_L= \sqrt{{\pi}/{2}}\omega_{pe}(r)\left({\rm v}/{\rm v}_{Te}\right)^3\exp\left(-{{\rm v}^2}/{2 {\rm v}_{Te}^2}\right)$ given by third RHS term of Equation (\ref{eqn:k2}), where ${\rm v}_{Te}=\sqrt{( k_BT_e/m_e)}$.
This latter term is included because we do not explicitly simulate the thermal plasma
electrons below approximately $3 {\rm v}_{Te}$ and their effect is therefore not accounted for by the first RHS terms.
Finally, we have two source terms, denoted $\mathrm{St}(W)$ and describing the scattering of Langmuir waves by plasma
ions and their interactions with ion-sound waves respectively, given in the next section.

\subsection{Non-linear Langmuir and ion-sound wave evolution}

The two terms  ${\mathrm St}_{Ion}$ and  ${\mathrm St}_{Decay}$ in Equation (\ref{eqn:k2}) describe the scattering off ions (nonlinear Landau damping) $L + i\rightleftarrows L' + i'$
and decay of Langmuir waves $L\rightleftarrows L' + s$, where $i,i'$ denote an initial and
scattered plasma ion and $s$ is an ion-sound wave.
The general expressions describing these processes  \citep[e.g.][]{1980MelroseBothVols, 1995lnlp.book.....T} have been written under the same one-dimensional approximation as in e.g. \citet{2002PhRvE..65f6408K}, which is that the fast-electron generated Langmuir waves propagate approximately parallel to the generating electrons and therefore the ambient magnetic field.
{In this case, both of these nonlinear processes produce back-scattered (negative wavenumber) Langmuir waves, approximately antiparallel to the initial Langmuir waves. The ion-sound waves produced are also approximately beam parallel or anti-parallel, and the dynamics remain  1-dimensional.
Hereafter, we omit the explicit time and space dependence of the spectral energy densities for clarity of notation.}
Hence the evolution of Langmuir wave spectrum due to scattering by ions is described by
\begin{align}\label{eqn:LIonA}
{\rm St}_{ion}(W(k))=\int dk_{L'} \frac{\alpha_{ion}}{|k-k_{L'}|} \exp{\left(-\frac{(\omega_L-\omega_{L'})^2}{2|k-k_{L'}|^2 {\mathrm v}_{Ti}^2}\right)} \times \notag
\\\left[\frac{1}{k_BT_i}\frac{\omega_{L'}-\omega_L}{\omega_L}W(k_{L'})W(k)\right]
\end{align}
where $k, k_{L'}, \omega_L, \omega_{L'}$ are the wavenumber and frequency of the initial and scattered Langmuir waves, and
\begin{equation}
\alpha_{ion}=\frac{\sqrt{2\pi} \omega_{pe}^2}{4 n_e {\mathrm v}_{Ti} (1+T_e/T_i)^2},
\end{equation} with ${\mathrm v}_{Ti}=\sqrt{k_B T_i/M_i}$ the ion thermal speed, and $M_i$ the mass of a plasma ion.
It is evident from the exponential factor that the scattering is strongest for $\omega_L \simeq \omega_{L'}$ and $k_{L'} \simeq -k$,
i.e. for backscattering of the waves. The resulting momentum change for the Langmuir wave is absorbed
by the ions which we assume to have a Maxwellian distribution at temperature $T_i$.
This momentum transfer is small, so the deviation of the ion distribution from thermal can be neglected \citep{1995lnlp.book.....T}.

The second source term describes Langmuir wave decay, and is given by
\begin{align}\label{eqn:4ql_sSrc}
&{\rm St}_{decay}(W(k))=\alpha_S\omega_{k} \int dk_S \omega_{k_S}^S \times \notag \\ &
\Biggl[ \left(
\frac{W(k_L)}{\omega^L_{k_L}}\frac{W_S(k_S)}{\omega^S_{k_S}}-
\frac{W(k)}{\omega^L_k}\left(\frac{W(k_L)}{\omega^L_{k_L}}+
\frac{W_S(k_S)}{\omega^S_{k_S}}\right)\right)\delta (\omega^L_{k}-\omega^L_{k_L}-\omega^S_{k_S})\Biggr.
\notag \\&
-\left.
\left(
\frac{W(k_{L'})}{\omega^L_{k_{L'}}}\frac{W_S(k_S)}{\omega^S_{k_S}}-
\frac{W(k)}{\omega^L_k}\left(\frac{W(k_{L'})}{\omega^L_{k_{L'}}}-
\frac{W_S(k_S)}{\omega^S_{k_s}}\right)\right)\right.\times \notag \\ & \Biggl.\delta(\omega^L_{k}-\omega^L_{k_{L'}}+\omega^S_{k_S}) \Biggr],
\end{align}
where $W_S(k_S), \omega_{k_S}^S$ are the spectral energy density and frequency of ion-sound waves, given by $\omega_{k_S}^S=k_S {\mathrm v}_s$ with ${\mathrm v}_s=\sqrt{k_BT_e(1+3T_i/T_e)/M_i}$ the sound speed, and the constant is
\begin{equation}
\alpha_S=\frac{\pi \omega^2_{pe}(1+3T_i/T_e)}{4n_ek_B T_e}.
\end{equation}

For a given initial Langmuir wavenumber, $k$, we have two possible processes, namely $L \rightarrow L' +s$ and  $L +s \rightarrow L'$. The wavenumbers of the resulting Langmuir wave, $k_L, k_{L'}$ respectively, and the participating ion-sound wave, $k_S$, are found from simultaneous solution of the equations of energy conservation (encoded by the delta functions in Equation \ref{eqn:4ql_sSrc}), and momentum conservation, given by $k_L= k-k_S$ and $k_{L'}=k+k_S$ for the two processes respectively. For example, for the process $L \rightarrow L' +s$ we find $k_{L} \simeq - k$, and $k_S \simeq 2 k$, and the initial Langmuir wave is backscattered. More precisely, we have $k_{L} = - k + \Delta k$ with the small increment $\Delta k= 2 \sqrt{m_e/M_i} \sqrt{(1+3T_i/T_e)}/(3\lambda_{De})$. Thus repeated scatterings tend to accumulate Langmuir waves at small wavenumbers.

Similarly, the evolution of the ion-sound wave distribution is given by
\begin{align}\label{eqn:4ql_s}
&\frac{\partial W_S(k)}{\partial t}=-\gamma_S(k)
W_S(k)\notag \\&-\alpha_S ({\omega^S_k})^2\int
\left(
\frac{W(k_L)}{\omega^L_{k_L}}\frac{W_S(k)}{\omega^S_{k}}-
\frac{W(k_{L'})}{\omega^L_{k_{L'}}}\left(\frac{W(k_L)}{\omega^L_{k_L}}+
\frac{W_S(k)}{\omega^S_{k}}\right)\right)\times\notag \\&
\delta(\omega^L_{k_{L\prime}}-\omega^L_{k_L}-\omega^S_k)dk_{L\prime}.
\end{align}
The second term here is analogous to Equation \ref{eqn:4ql_sSrc}, describing the {interaction of an ion-sound wave at wavenumber $k$ with a Langmuir wave at wavenumber $k_L$, producing a Langmuir wave at wavenumber $k_{L'}$.} {Again, these participating wavenumbers are found from simultaneous solution of energy (frequency) and momentum (wavenumber) conservation.} The first term is Landau damping of the waves,
with coefficient
\begin{equation}
\gamma_S(k)=\sqrt{\frac{\pi}{2}}\omega^S_k\left[\frac{{\mathrm v}_s}{{\mathrm v}_{Te}}+\left(\frac{{\mathrm v}_s}{{\mathrm v}_{Ti}}\right)^3\exp\left[- \left(\frac{{\mathrm v}_s}{2{\mathrm v}_{Ti} }\right)^2 \right]\right].
\end{equation}

\subsection{Electromagnetic emission}
Electromagnetic emission is described in terms of its brightness temperature,
$T_T$, which is defined from the Rayleigh-Jeans law for the radiation intensity
as function of frequency $f=\omega/(2\pi)$
by
\begin{equation}\label{eqn:RJ}
I(f)=2 f^2  k_B T_T/c^2.
\end{equation}

As stated above, the electrons, Langmuir waves, and ion-sound waves are treated one-dimensionally, along the ambient magnetic field. Although the electromagnetic emission pattern produced from Langmuir waves is  azimuthally symmetric around the beam direction, it is not aligned along the beam and requires a different approach. Radio emission near the plasma frequency has a dipole pattern with a peak at polar angle $\pi/2$ to the direction of the beam, while electromagnetic waves produced near double the plasma frequency (the harmonic) peak at $\pi/4$ and $3\pi/4$ radians \citep{1970SvA....14..250Z}. {Because our model assumes azimuthal symmetry around the beam direction, we do not distinguish the lobes at $+\pi/2$ and at $-\pi/2$, and similarly those at $\pm\pi/4$ and $\pm3\pi/4$. }
Here, we use an angle-averaged model as described in previous work \citep{2014A&A...562A..57R}, by averaging the emission probability over angles using the assumed small angular spread of the Langmuir waves, and including the constraints on the wavevectors involved as described in the following sections. The resulting angle-averaged electromagnetic spectral energy density depends only on the magnitude of the wavevector, and so the derived brightness temperature is also angle independent.  We do not assume that the harmonic wavenumber is far smaller than the Langmuir wavenumber, as is often done in the so-called head-on approximation
\citep{1979A&A....73..151M}.

For thermal radiation (free-free emission), the brightness temperature is $T_T=T_e$, where $T_e$ is the plasma temperature.
From Kirchoff's law the thermal emission rate $P(k) = \gamma_d T_e$
is related to the damping rate $\gamma_d$ giving the thermal
radiation level.
Thus, from the bremsstrahlung damping rate,
\begin{equation}\label{eqn:gammaD}
\gamma_{d}(k)=\gamma_{c}\frac{\omega_{pe}^2}{\omega(k)^2},
\end{equation}
one can find the thermal emission rate.

\subsection{Escape of electromagnetic emission}
To treat the propagation of radiation between the source and observer, we assume the simple case of travel approximately along the ambient magnetic field, and thus along the plasma density gradient. {At these distances ($<10 R_{sun}$)  this is approximately parallel to the plasma density gradient} \citep[e.g.][]{1958ApJ...128..664P}. The equation describing electromagnetic radiation transfer in inhomogeneous, magnetised plasma along the direction of the magnetic field and inhomogeneity is \citep[][Equation 10]{1969ApJ...155.1129Z}
\begin{equation}\label{eqn:Zhelez}
\frac{1}{{\rm v}_{g}} \frac{\partial I(f) }{\partial t} + k^2 \frac{d}{dl}\left(\frac{I(f) }{k^2} \right) = a(f) - \mu I(f)
\end{equation}
{where $l$ is the path length along a ray}, $I(f)$ is the spectral intensity, ${\rm v}_{g}=\partial \omega/\partial k$ is the group velocity of electromagnetic waves with the dispersion relation $\omega^2(k)=\omega_{pe}^2+c^2k^2$.
$a(f)$ is the radiation source and $\mu$ is the absorption coefficient.
The absorption is assumed to be due to collisions, so that $\mu =\gamma_d/{\rm v}_{g}$.
The emission $a(f)$ includes the thermal radiation and the coherent
plasma radiation by Langmuir waves (noting that the latter depends
on radiation intensity $I(f)$.)

Using the Rayleigh-Jeans law (Equation \ref{eqn:RJ}) to relate the spectral intensity
and the radiation brightness temperature,
and expanding $d/dl$ \citep[][Eq 4.28 with Eq 4.23]{1996ASSL..204.....Z},
we obtain the equation for the evolution of radiation brightness temperature namely
\begin{equation}
\label{eqn:EMEvol}
\frac{\partial }{\partial t}T_T(k) + \frac{\partial \omega}{\partial k} \frac{\partial}{\partial r}T_T(k) - \frac{\partial \omega}{\partial r} \frac{\partial}{\partial k}T_T(k) = \gamma_d T_e - \gamma_{d} T_T(k) + \mathrm{S.\,T.},
\end{equation}
{where $r$ is the direction of beam propagation}, the first term on the RHS is the thermal emission rate and the second term is the {collisional} absorption.
$\mathrm{S.\,T.}$ are the source terms describing the production of electromagnetic waves via
the non-linear processes described in the following subsections.

\subsection{Fundamental electromagnetic source terms}\label{sec:Fund}

The processes for emission at the fundamental are $L\rightleftarrows t \pm s$ where $L,s$
are Langmuir and ion-sound waves, and $t$ is an electromagnetic (EM) wave,
and $L + i\rightleftarrows t + i'$, for $i,i'$ an initial and final plasma ion.
The probability of both processes \citep[e.g.][]{1995lnlp.book.....T} has a maximum when the wavevector of the EM wave is
perpendicular to the initial Langmuir wave (dipole emission peaking at $\pi/2$ rad).

{The latter process, $L + i\rightleftarrows t + i'$,  is analogous to that described by Equation (\ref{eqn:LIonA}) for the conversion of Langmuir waves into EM waves}. However, {the growth rate for this process is much lower than that for the ion-sound wave interaction,} and for the cases considered here {is expected to be negligible} in comparison to this {\citep[e.g.][]{2000PhPl....7.4901C}.} Therefore we do not consider {direct ion scattering in this work.}

{For EM emission by the process $L\rightleftarrows t \pm s$ to be efficient, we require ion-sound waves to be present. These can be generated by the decay of Langmuir waves described above. It is important that these ion-sound waves, and the Langmuir waves, are not confined exactly to angles parallel to the electron beam, as the emission probability is proportional to $|\vec{k} \times \vec{k}_T|$ for $\vec{k}$ the wavenumber of the participating ion-sound wave and $\vec{k}_T$ the EM wave. {We therefore assume that the beam-generated Langmuir waves have some small angular spread in wavenumber space, covering a solid angle of $\Delta \Omega$. Further assuming that this solid angle forms a cone around the beam direction, we require it has a half-angle of the order 10 degrees for almost all waves to satisfy the kinematic conditions for interactions. For beam-generated Langmuir waves, backscattered Langmuir waves and the generated ion-sound waves, this condition is easily satisfied.} For example, 2D simulations of beam-plasma
relaxation \citep{1980SvJPP...6..232C,1989FizPl..15..790D,2008PPCF...50h5011Z} suggest that the beam remains relatively narrow and the magnetic field tends to support beam-plasma interaction in 1D \citep{1994PhPl....1.1821D}}.

{Further assuming the Langmuir waves are uniform within this angle, the resulting EM emission may be assumed approximately isotropic. Then using the general expressions in} \citep[e.g.][]{1980MelroseBothVols, 1995lnlp.book.....T}, {averaging over angles and rewriting in terms of the} hemisphere-averaged brightness temperature, {given by} \begin{equation}\label{eqn:tbW} T_T(k_T)=\frac{W_T(k_T)}{2 \pi k_B k_T^2}
\end{equation} {for $W_T(k_T)$ the EM wave spectral energy density,} we obtain the expression given in \citet{2014A&A...562A..57R}:
\begin{align}
\label{eqn:FundS}
&\mathrm{St}_{fund}^{lts}(T_T(k))=
\frac{\pi \omega_{pe}^4{\mathrm v}_s\left(1+\frac{3T_i}{T_e}\right)}{ 24 {\mathrm v}_{Te}^2n_eT_e} \times \int dk_L  \notag \\
&\left\{ \left[\frac{W_S(k_S)}{\omega_{k_S}^S} \frac{2\pi^2}{ k_B k_L^2\Delta\Omega} \frac{W_L({k}_L)}{\omega_{k_L}^L}-\frac{T_T({k})}{\omega_{k}^T}\left( \frac{ W_S(k_S)}{\omega_{k_S}^S} + \frac{W_L(k_L)}{\omega_{k_L}^L}\right)\right]\right. \times \notag \\ &\delta(\omega_{k_L}^L +\omega_{k_S}^S-\omega_{k}^T)   \notag\\\
&+\left.\left[\frac{W_S(k_{S'})}{\omega_{k_{S'}}^S} \frac{2\pi^2}{k_B k_L^2\Delta\Omega} \frac{W_L({k}_L)}{\omega_{k_L}^L}-\frac{T_T({k})}{\omega_{k}^T}\left( \frac{ W_S(k_{S'})}{\omega_{k_{S'}}^S} - \frac{W_L(k_L)}{\omega_{k_L}^L}\right)\right]\right.\times\notag \\&\left. \delta(\omega_{k_L}^L -\omega_{k_{S'}}^S-\omega_{k}^T)\right\}. \end{align}
where the participating wavenumbers are obtained from energy and momentum conservation{(equivalent to frequency and wavenumber conservation)}. However, in this case the momentum conservation condition must be obtained from the 3-D description of the process. Using our assumption that the EM emission occurs approximately perpendicular to the initial Langmuir wave, the wavenumbers $k_L, k_S, k_T$ form a right-triangle and thus we find that $k_S^2=k_T^2 + k_L^2$.


\subsection{Harmonic electromagnetic emission source terms}\label{sec:Harm}

{Emission at the harmonic of the plasma frequency occurs due to the coalescence of two Langmuir waves, $L + L' \rightleftarrows t$. The probability peaks when the angle between initial Langmuir and final EM wavevectors is $\pi/4$ or $3\pi/4$, noting that the whole problem has azimuthal symmetry around the beam direction. However, the probability depends also on the magnitudes of the wavenumbers $k_1, k_2$ for the participating Langmuir
waves, and $k_T$ for the EM wave.} {Rather than assume head-on coalescence, we instead assume that emission occurs primarily at the angle where the probability is maximised, which is close to a $\pi/4$ or $3\pi/4$ angle between the initial Langmuir and final EM wavevectors. Using this assumption we use the wavevector and frequency matching to calculate an emission probability.}

{Writing $k_1, k_2$ for the wavenumbers of the two coalescing Langmuir waves, $\omega_{k_1}, \omega_{k_2}$ the corresponding Langmuir wave frequencies, and $k, \omega^T_{k}$ the wavenumber and frequency of the EM wave, we solve the energy and momentum conservation equations, given by $\omega_{k_1}+\omega_{k_2}=\omega_{k_T}$
and $\vec{k}_1+\vec{k}_2=\vec{k}$ respectively, we then obtain}
\begin{equation}k_1\simeq\frac{1}{2}k \cos\left(\frac{\pi}{4}\right) +\frac{1}{2} \sqrt{4 \frac{\omega_{pe}(\omega^T_{k} -2\omega_{pe})}{3 {\mathrm v}_{Te}^2} + k^2 \left(\cos^2\left(\frac{\pi}{4}\right)-2\right)},\end{equation}
and $k_2^2=k_1^2 +k^2-2k_1k \cos(\pi/4)$.

{We then average the emission probability, given by the general expressions of e.g. } \cite{1980MelroseBothVols, 1995lnlp.book.....T}, {over angles assuming the same geometry, and convert to brightness temperature using Equation} (\ref{eqn:tbW}). The resulting source term for harmonic emission is, as in previous work \citep{2014A&A...562A..57R}: \begin{align}&\mathrm{St}_{harm}^{ll't}(T_T(k)) =\omega_{k}^T \frac{\pi \omega_{pe}^2}{48 m_e n_e {\mathrm v}_{Te}^2} \int d k_1 \frac{(k_2^2-k_1^2)^2}{4 k_2^2} \times \notag
\\ &  \left[\frac{2\pi^2}{k_B k_2^2 \Delta\Omega}\frac{W(k_1)}{\omega_{k_1^L}}\frac{W(k_2)}{\omega_{k_2}^L}-\frac{T_T(k)}{\omega_{k}^T} \left(\frac{W(k_1)}{\omega_{k_1}^L}+\frac{W(k_2)}{\omega_{k_2}^L}\right)\right]\times \notag \\& \delta(\omega_{k_1}+\omega_{k_2}-\omega_{k}^T).
\end{align}

\subsection{Observed radiation fluxes}

Because Langmuir waves are present over a range of wavenumbers, the emission at a given frequency arises from a range of spatial locations. Since the waves propagate at constant frequency, as they travel into lower density plasma their wavenumber increases. After a short distance, the wavenumber becomes too large to satisfy the momentum and energy matching conditions for interaction with Langmuir waves. We model the propagation of radiation using Equation (\ref{eqn:Zhelez}) through this interaction region.

After the radiation can no longer interact with Langmuir waves, we consider it as propagating directly from source to observer, along the ambient magnetic field, without angular scattering. Absorption due to inverse bremsstrahlung is included using Eq. (\ref{eqn:gammaD}), as this can be significant at the highest frequencies in the simulation. The resulting optical depth is given below. {We assume the ambient magnetic field expands in a cone, and thus the cross sectional area of the electron beam expands similarly. We assume this cross section is circular and that emission is produced equally over the entire area. Thus we model the emission source as a circle subtending a solid angle $\pi \theta^2$ where $\theta$ is its half-angular size as seen by the remote observer. Further, we currently assume the entire source is visible to the remote observer, and that emission directionality is unimportant.} The dependence of EM group velocity on wavenumber
$$
{\rm v}^T_g =\frac{\partial\omega}{\partial k_T}=\frac{c^2k_T}{\sqrt{\omega_{pe}^2+c^2k_T^2}}
$$
produces time delays between the arrival of the fundamental
and harmonic components, and between radiation from different locations. These delays between the {source at $r_{Src}$ and observer at $r_{Obs}$} are accounted for
by integrating group velocity over distance
$$
t_{Delay}=\int_{r_{Src}}^{r_{Obs}} \left(\frac{1}{{\mathrm v}_g^T(r)} -\frac{1}{c} \right)dr,
$$
and are included in the dynamic spectra shown below.

{Changing Equation (\ref{eqn:RJ}) from intensity to flux as function of frequency, defined by $F(f)= I(f) \pi \theta^2$, where $\theta$ is the source angular size}, the observed flux density at the Earth for radiation with in-source brightness temperature
$T_T$ is
\begin{equation}\label{eqn:Flux}
F(f)= 2 k_B T_T \frac{f^2}{c^2} \pi \theta^2 \exp{(-\tau)}.
\end{equation}
Here $\tau$ is the optical depth for propagation from source to observer due to inverse bremsstrahlung absorption. We calculate this assuming a locally exponential background density profile with scale height $H$, i.e. $n_e(r) \propto \exp(- (r-r_{Src})/H)$. The actual density used in the code is given by Equation (\ref{sol1}), but the locally exponential model suffices for this simple optical depth calculation. We obtain \begin{equation}\label{eqn:optDep}
\tau=\int_{r_{Src}}^{r_{Obs}} \frac{\gamma_d(r)}{{\mathrm v}_g^T(r)} dr.
\end{equation} Integration
of Equation (\ref{eqn:gammaD}) gives
\begin{align}\label{eqn:optDepFin}
&\tau=\sqrt{\frac{2}{\pi}}\frac{e^2 \ln\Lambda}{12 \pi^2 {\mathrm v}_{Te}^3 m_e}\frac{H}{ c}\times \notag\\&\left[ f_0^2 - \frac{1}{f_0}\sqrt{(f_0^2-f_{pe}^2(0))}\left(f_0^2 +0.5f_{pe}^2(0) \right) \right],
\end{align}
where $f_0$ is the frequency of the emission and $f_{pe}(r)$ the local plasma frequency {at position $r$}. In general $\exp{(-\tau)}$ is of order unity
for the frequencies considered here, although for fundamental emission near 500~MHz it falls as low as 0.1.

\begin{figure*}
\centering
\sidecaption
\includegraphics[width=12cm]{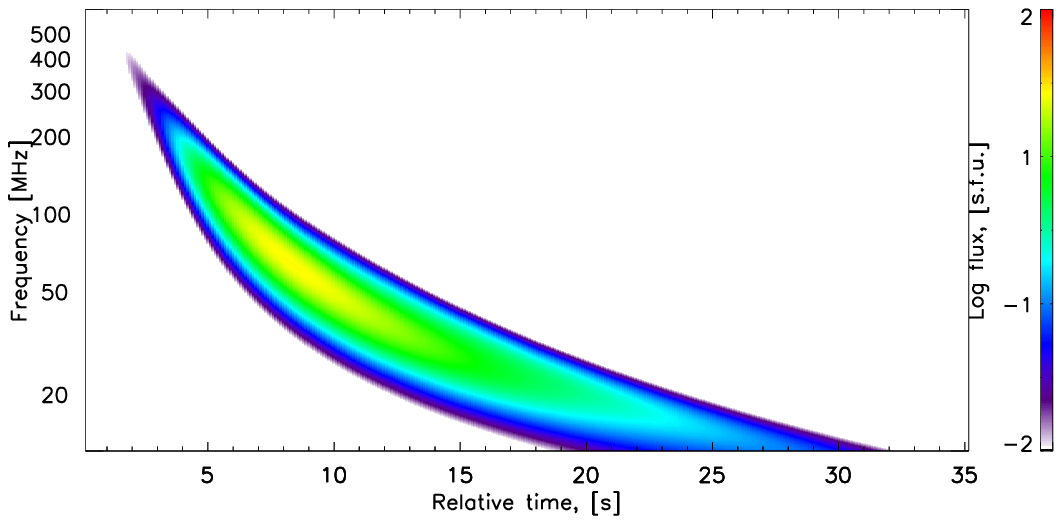} \caption{The dynamic spectrum of simulated type III radio burst emission (flux at 1AU)
for almost instantaneous injection of energetic electrons around 550~MHz. 
{Differences in propagation time from the changing source position to the remote observer are included and the time axis is set such that emission from the injection region would appear at $t=0$.}
}
\label{fig:DynSpecHom}

\end{figure*}

\section{Initial conditions} \label{sec:InitCond}

We model the injection of energetic electrons via the source term added
to Equation (\ref{eqn:k1}) with separate functions describing velocity, space and time dependencies
\begin{equation}\label{eq:init_f}
S({\rm v},r,t) = g_0({\rm v})h_0(r)i_0(t).
\end{equation}
Based on the simultaneous observations of energetic electrons at the Sun and near the Earth \citep[e.g.][]{2007ApJ...663L.109K}
as well as the type III and  X-ray emissions \citep[e.g.][]{2005SoPh..231..117A,2011A&A...529A..66R},
we assume a power-law distribution in velocity:
\begin{equation}
g_0({\rm v}) = \frac{n_{b}(2\delta -1) }{{\rm v}_{\rm{min}}}\left(\frac{{\rm v}_{\rm{min}}}{{\rm v}}\right)^{2\delta}
\end{equation}
where $\delta$ is the spectral index of the injected electrons in energy space (with $\delta =3$ for the simulations),
and $n_b$ is time integrated beam number density. ${\rm v}_{\rm{min}}$ is the minimum beam velocity
which is set at $3~{\rm v}_{Te}$, below which Langmuir waves are
heavily Landau damped and can be safely neglected. The maximum velocity simulated is set to $2\times10^{10}~\rm{cm~s}^{-1}$ to avoid relativistic effects which are not accounted for. Because of the power-law form of the injected electron distribution, the number of electrons above this velocity is negligibly small.
The injection time profile is assumed to be an asymmetric Gaussian:
\begin{equation}
i_0(t)=\frac{2}{\sqrt{\pi}(\tau_1+\tau_2)}\exp\left(-\frac{(t-t_0)^2}{\tau^2}\right).
\label{eqn:time_inj}
\end{equation} {with separate timescales, so that $\tau=\tau_1$ during the rise
at times $t\le t_0$, and $\tau=\tau_2$ during decay at $t>t_0$}:
We set $t_0=4\tau_1$, allowing time for the rise phase. In the simulations below we consider
both almost instantaneous injection by setting $\tau_1=\tau_2 =10^{-3}$~s, and slower injection with $\tau_1=1$~s and $\tau_2=4$~s.
This value is sufficiently long to show the effects of injection on the observed emission,
while being small enough that computational time is reasonable.
Finally, the spatial distribution of the electron injection is also Gaussian, with
\begin{equation}
h_0(r)=\exp\left(-\frac{(r-r_i)^2}{d^2}\right)
\label{eqn:space_inj}
\end{equation}
where $r_i$ is the location of the injection region, and $d$ its spatial size. We set $r_i=2\times 10^9$~cm, corresponding to a local plasma frequency of approximately 550~MHz, and $d=10^9$~cm. The cross-sectional radius of the source in the injection region is taken as $2\times 10^9$cm. Radial expansion of the magnetic field as the beam propagates
will lead to a steadily increasing source size, giving angular sizes of $0.8^\prime$ at 432~MHz, $2.5^\prime$ at 169~MHz, and $6.3^\prime$ at 43~MHz, comparable to observed type III source sizes \citep[e.g.][]{1985srph.book.....M,2013ApJ...762...60S}.

%

The initial Langmuir wave spectral energy density and radiation brightness temperatures are set to the thermal level:
\begin{equation}\label{eqn:LTh}
W(x,k, t=0)= \frac{k_b T_e}{4 \pi^2}k^2\ln\left(\frac{1}{k\lambda_{De}}\right),
\end{equation} with $\lambda_{De}=1/k_{De}={\mathrm v}_{Te}/\omega_{pe}$
and $T_T=T_e$. The three-wave decay processes and ion-sound wave damping rapidly
establish the level of ion-sound waves, and so we set the initial level
of $W_s$ to a small, non-zero value.

{The background plasma is assumed to be isothermal at 1.5MK, and the electron and ion temperatures are set everywhere equal. }
The plasma density profile is defined as in \citet{2013SoPh..285..217R}, based on the model of \citet{1958ApJ...128..664P}
with normalisation as in \citet{1999A&A...348..614M}, as the solution of
\begin{equation}\label{sol1}
r^2n_0(r)u(r)= C= const
\end{equation}
where $n_0$ is the plasma density, $u(r)$ solar wind speed and
\begin{equation}\label{sol2}
  \frac{u(r)^2}{u_c^2}-\mbox{ln}\left(\frac{u(r)^2}{u_c^2}\right)=
  4\mbox{ln}\left(\frac{r}{r_c}\right)+4\frac{r_c}{r}-3
\end{equation}
where $v_c\equiv u(r_c)=(k_BT_e/\tilde{\mu}m_p)^{1/2}$, $r_c=GM_s/2v_c^2$, $T_e$ is the electron temperature,
$M_s$ is the mass of the Sun, $m_p$ is the proton mass, $\tilde{\mu}$ is the mean molecular weight and the constant
is $C=6.3\times 10^{34}$ s$^{-1}$.

\section{Numerical Results}

\subsection{Instant electron injection}

As an initial example, we consider electrons injected almost instantaneously ($\tau_1=\tau_2=10^{-3}$~s) at a height of $2\times 10^9$~cm, with source function given by Equation \ref{eq:init_f} with $n_b=4\times 10^6$~cm$^{-3}$.
The resulting dynamic spectrum is shown in Figure \ref{fig:DynSpecHom}. The injected power law of electrons is initially stable to Langmuir wave generation and so no emission is produced at the injection site. After a certain distance \citep[e.g.][]{2011A&A...529A..66R} the fast electrons outpace the slower ones and a reverse slope in velocity space is formed,
 leading to Langmuir wave generation
and radio emission. The onset frequency therefore corresponds to a plasma frequency only slightly smaller
than that of the injection site. {The strong long-duration component in the spectrum in Figure \ref{fig:DynSpecHom} is harmonic ($2 f_p$) emission, with  the fundamental fluxes staying below 1~sfu. This emission therefore arises from a region where the plasma frequency is about half that of the emission itself.}
For comparison, the quiet Sun flux from the whole sun is, using average values from \citet{2009LanB...4B..103Ba}, around $27$~sfu at $500$~MHz, $8$~sfu at $200$~MHz and $0.07$~sfu at $20$~MHz.

\begin{figure}[h]
\centering \includegraphics[width=0.4\textwidth]{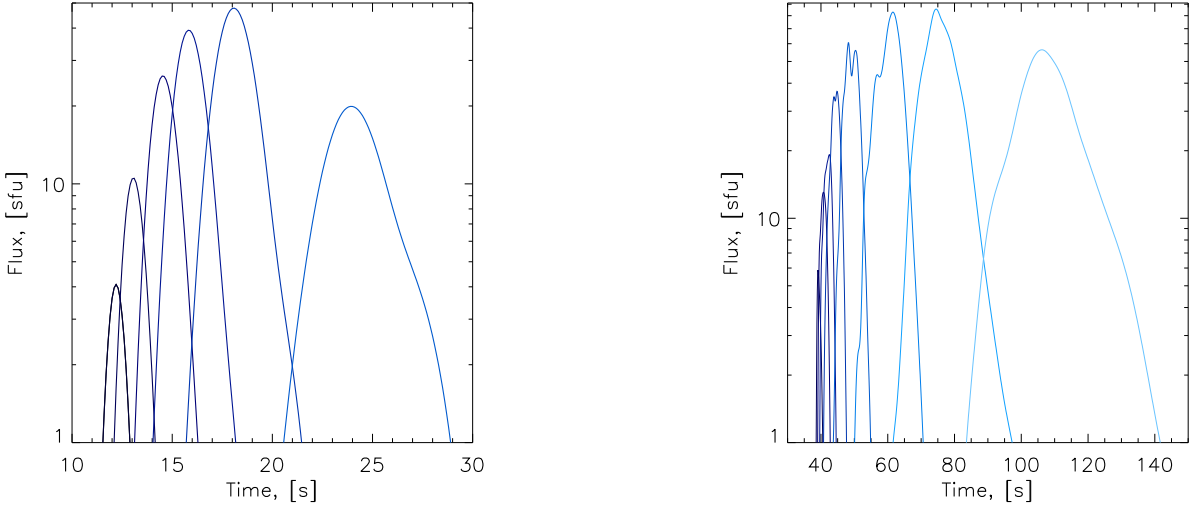}
\caption{Radio emission time profiles at the Earth for nearly instant electron injection, at frequencies of 200, 150, 100, 75, 50 and 25 MHz (dark to light blue respectively).}
\label{fig:profinst}
\end{figure}

Time profiles of the emission are shown in Figure \ref{fig:profinst} for frequencies of 200, 150, 100, 75, 50 and 25 MHz. The timescale for Langmuir wave damping, originally thought to dictate the duration of Type III bursts, increases as frequency decreases. {This is consistent with the observations by e.g.} \citet{1972A&A....19..343A,1973SoPh...31..501E,1987A&A...175..271S}. {These considered emission at frequencies from 100s of MHz to the low kHz, and found an approximate $1/f$ scaling of duration with frequency.} Our results however suggest it is not in fact this damping time which dictates the duration: this is discussed further in Section \ref{sec:duration} below. 

The time profiles show a slight asymmetry, with faster rise and slower decay, and this asymmetry becomes stronger as frequency increases. For example, at 200~MHz the ratio of decay time (half-width half-maximum) and rise time is 1, while at 25~MHz it is approximately 1.6. This is because the excitation of Langmuir waves by fast electrons continues alongside their decay, as the exciter has a finite spatial length. Moreover, this length is seen to increase with as frequency decreases, leading to a corresponding increase in the duration of Langmuir wave excitation, and thus the increasing rise-time of the emission.

\subsection{Slow electron Injection}

{As noted in the previous section,} the injected electrons have a power-law distribution in velocity space, which only becomes unstable to Langmuir wave generation
due to time-of-flight effects (producing the bump-on-tail instability). For an electron injection function which is a function of time, the time-of-flight effects are altered and therefore so are the development of the bump-in-tail instability and consequently the radio emission. {The faster electrons must now outpace both the slower electrons injected at the same time, and those injected earlier, before an unstable velocity space gradient can develop, leading to a decreased onset frequency of emission}. For the remainder of this paper we consider electrons injected with $\tau_1=1$~second and $\tau_2=4$~seconds. We take a time-integrated beam density of $n_b=5\times 10^6$~cm$^{-3}$. The total number of injected electrons above 50~keV, for cross-sectional source size $2\times10^9$~cm is then $10^{30}$ electrons. This is slightly below the lower limit of the observed densities \citep[e.g.][]{2007ApJ...663L.109K}, but this is compensated for by the short injection time chosen here for computational purposes, which leads to larger instantaneous electron densities.

\begin{figure*}
\sidecaption
\includegraphics[width=12cm]{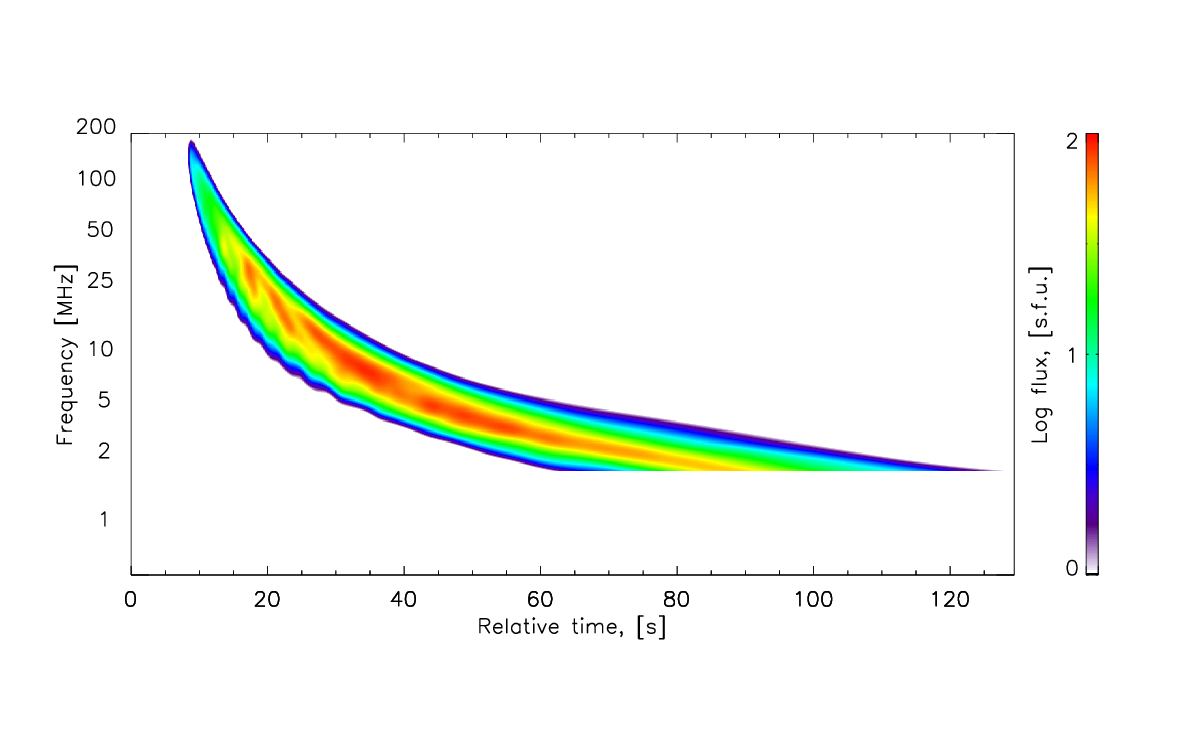}
\caption{The dynamic spectrum of a simulated type III radio burst for slow ($\tau_1=1$~second)
injection of energetic electrons around 550~MHz. {As in Figure \ref{fig:DynSpecHom} the time axis is set such that emission from the injection site would appear at $t=0$. }
}
\label{fig:DynSpecInj}

\end{figure*}

Figure \ref{fig:DynSpecInj} shows the dynamic spectrum for this case.
The later onset of the emission is evident compared to Figure \ref{fig:DynSpecHom}, due to the slower development of the bump-on-tail instability. The fundamental component is again weak, in this case reaching only about 0.4~sfu. {The fine structure in the dynamic spectrum is not a numerical effect, but probably related to the details of Langmuir wave scattering.}

\subsection{Burst duration, rise and decay times at a given frequency}
\label{sec:duration}
\begin{figure}
\centering \includegraphics[width=0.4\textwidth]{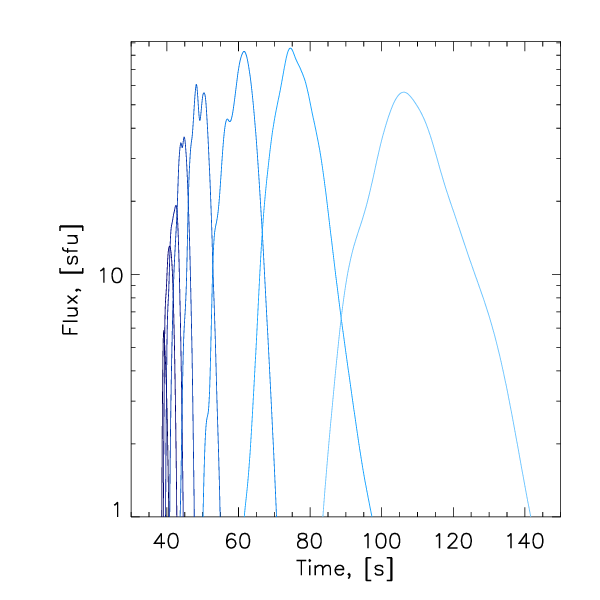} \\\includegraphics[width=0.4\textwidth]{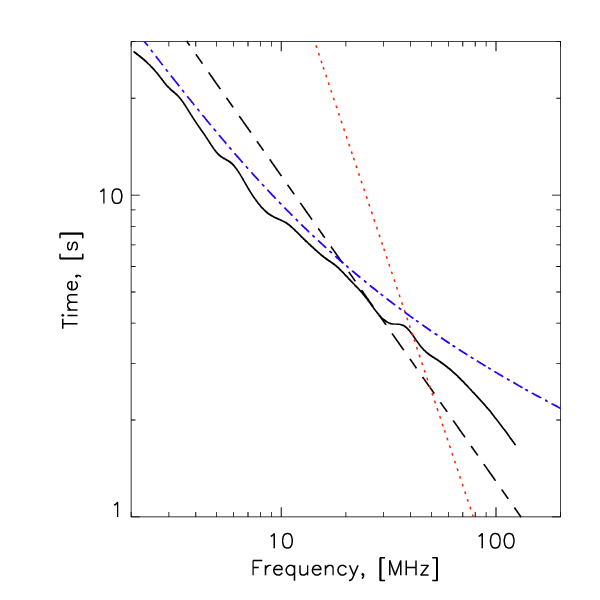}
\includegraphics[width=0.4\textwidth]{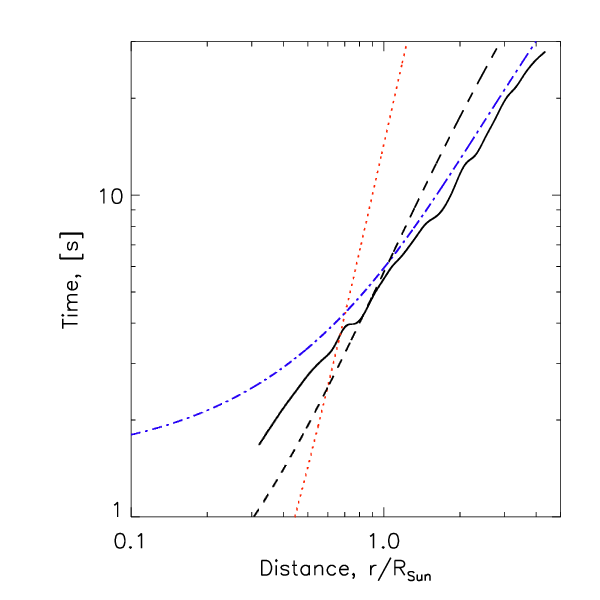}

\caption{Top: Radio emission time profiles at frequencies of 200, 150, 100, 75, 50, 25, 10 and 5 MHz (dark to light blue respectively)
for a 1~second rise time of electron injection. Middle/Bottom: Half-width-half-maximum of emission (black line), the collisional Langmuir wave decay time (Equation \ref{eqn:tcoll}, red dotted line), the timescale for Langmuir wave evolution due to plasma density inhomogeneity (Equation \ref{eqn:tinhom}, blue dot-dashed line) and the empirical result given by Equation \ref{eqn:dur} (black dashed line), {as function of both background plasma frequency and distance from injection site.}}
\label{fig:profinj}
\end{figure}

{Figure \ref{fig:profinj} shows time profiles of the emission at frequencies of 200, 150, 100, 75, 50, 25, 10 and 5 MHz. The simulations show fast exponential rise and somewhat slower exponential decay. The characteristic duration of the burst at a given frequency is then the sum of these two times.} 
In Figure \ref{fig:profinj} we plot the {half-width-half-maximum (HWHM)} duration of the emission,
calculated by finding the variance of the time profiles at each frequency. {The profiles are very close to symmetric and so this is approximately equal to both the rise and decay times.} For comparison, we plot the observationally derived empirical relationship of e.g. \citet{1973SoPh...30..175A},
 which gives a decay time
 \begin{equation}
 \label{eqn:dur} \tau_{Decay}= 10^{7.71}(10^6 f)^{-0.95}
 \end{equation}
 where $f$ is measured in MHz. {Similar $f^{-1}$ scalings were found by e.g} \citet{1972A&A....19..343A,1987A&A...175..271S}. We note that such empirical results are derived from a large number of bursts with varying exciter parameters, such as the injection timescale for accelerated electrons, injection region size, height etc. While our simulations use parameters within the inferred ranges, the typical values, and therefore those that would dominate the empirical result, are not known. Even the background plasma density profile chosen, while a reasonable assumption, may not fully reflect the real values. However, as is seen in Figure \ref{fig:profinj}, the simulated and empirical results agree well over the frequencies considered, always to within a factor of 2.

In order to discern the dominant physical factor in the decay time, in Figure \ref{fig:profinj} we also plot the collisional decay timescale,
given by
\begin{equation}
\label{eqn:tcoll}
\tau_{coll}\simeq \frac{m_e^2 v_{Te}^3} {\pi n_e e^4\ln\Lambda}
\end{equation}
 and that for spectral evolution of Langmuir waves due to density inhomogeneities, given by
 \begin{equation}\label{eqn:tinhom}
 \tau_{inhom} \simeq \frac{k}{(d\omega_{pe}/dx)}
 \end{equation}
 for typical wavenumber $k\sim 0.1 k_{De}$ for beam-generated Langmuir waves.

{It was originally suggested that the decay of type III bursts was due to collisional {damping} of Langmuir
waves \citep[e.g.][]{1963ApJ...138..239H,1972A&A....16....1E}. However, it is clear from this figure that the collisional decay of Langmuir waves is not the dominant factor.} At high frequencies, it is far shorter than the decay time. The rise and decay times are similar (seen in the time profiles in Figure \ref{fig:profinj}) and the duration of excitation appears to be the main factor. At lower frequencies, the collisional time is far longer than the burst, and the burst decay time is longer than the rise time, although in this case only slightly. Here it appears the density inhomogeneity is responsible for the decay, as shown by the similar frequency scalings of the two. Finally we note that there is a region near 40~MHz (for these parameters) where the collisional timescale and the burst decay times are quantitatively similar. This may explain both the apparent successes and errors of previous attempts to use decay time to derive plasma temperature \citep[e.g.][]{1973SoPh...30..175A,1974A&A....31..419B,1974SoPh...34..181R}.

\begin{figure}
\centering \includegraphics[width=0.4\textwidth]{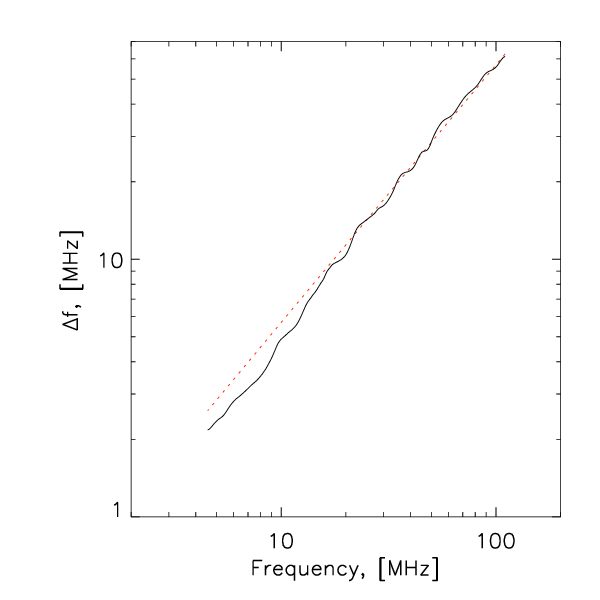} \includegraphics[width=0.4\textwidth]{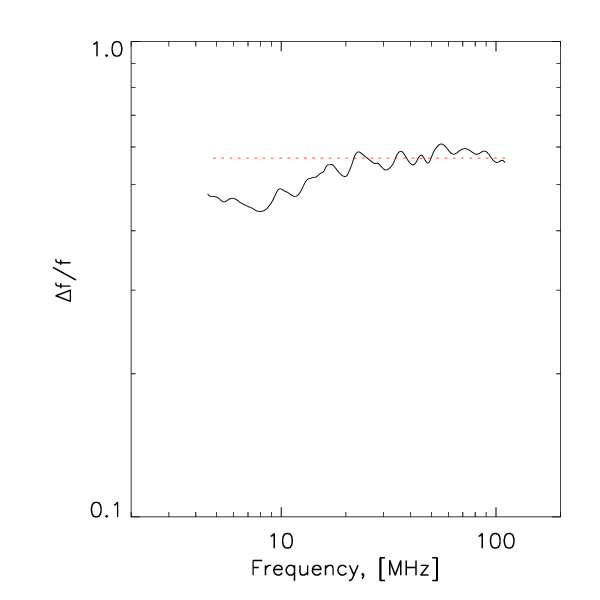}
\caption{Top: Full Width at Half Maximum (FWHM) bandwidth of radio emission flux at a given time as a function of the peak emission
frequency. Bottom: the bandwidth divided by the peak frequency, as a function of the peak frequency. The red dashed lines are a linear fit, $\Delta f = 0.57 f$.}
\label{fig:Bandwidth}
\end{figure}

\subsection{Instantaneous bandwidth}
{ In Figure \ref{fig:Bandwidth} we plot the instantaneous bandwidth of the emission as a function of frequency, that is the FWHM bandwidth at a given time}. The measured bandwidth is a significant fraction of the frequency, from 0.4 to 0.6 here.
The frequency spread of Langmuir waves at a single location is small, with wavenumber spread $k\lambda_{De} \sim 0.2$, varying very little with location, and corresponding to $\Delta f /f = 0.06$. To obtain a value of say $\Delta f/f = 0.5$ 
 from a single location would require a much larger wavenumber range of $k\lambda_{De} \sim 0.6$ 
  as noted by \citet{2011SoPh..269..335M}. On the other hand, at a given instant Langmuir waves are present over a significant spatial length. This spread in space reflects the spatial extent of the fast electrons, which increases over time as the slower electrons are left behind by faster ones, and is easily able to explain the observed bandwidths.

\subsection{Frequency drift}
\begin{figure}
\centering
\includegraphics[width=0.4\textwidth]{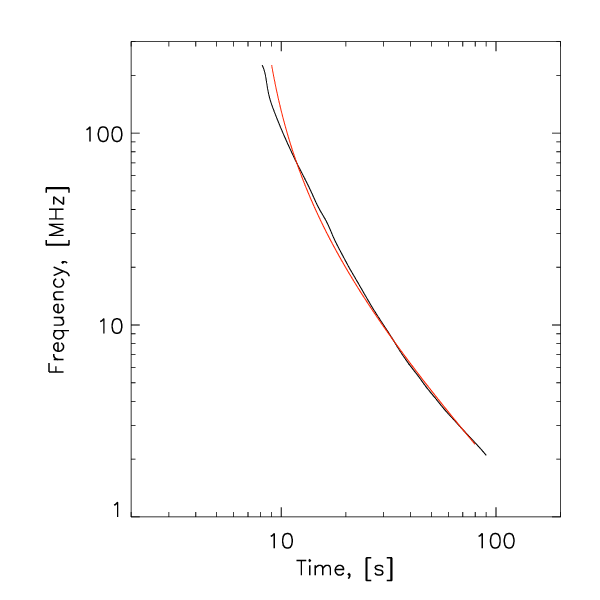}
\includegraphics[width=0.4\textwidth]{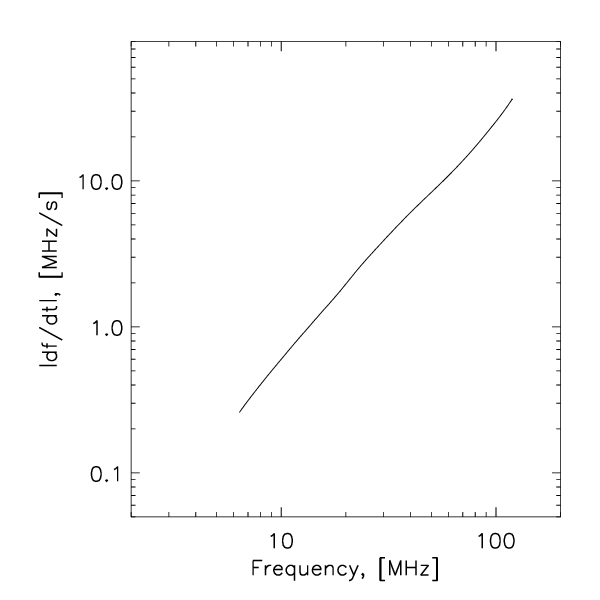}

\caption{Top: The peak emission frequency as a function of time. The red line is a single power-law fit of the form $f=a(t-t_0)^b$, where $t_0$ accounts for the onset time of emission and we have $a\simeq 150$, $b\simeq -0.6$. Bottom: the frequency drift rate, found by fitting power-laws to segments of the peak emission curve and differentiating these, as described in the text.}
\label{fig:drift}
\end{figure}

{The frequency drift of the burst is related, {although not straightforwardly}, to the speed of the exciting electrons, and thus {should} show which energy electrons are most important for emission. }
To find this, we first find the peak emission frequency as a function of time. This is plotted in the left panel of Figure \ref{fig:drift}, along with a power law fit given by the red line. Using this fit, which gives $f(t)$, we may analytically differentiate and rearrange to find $df/dt$
as a function of $f$.

\begin{figure}
\centering
\includegraphics[width=0.4\textwidth]{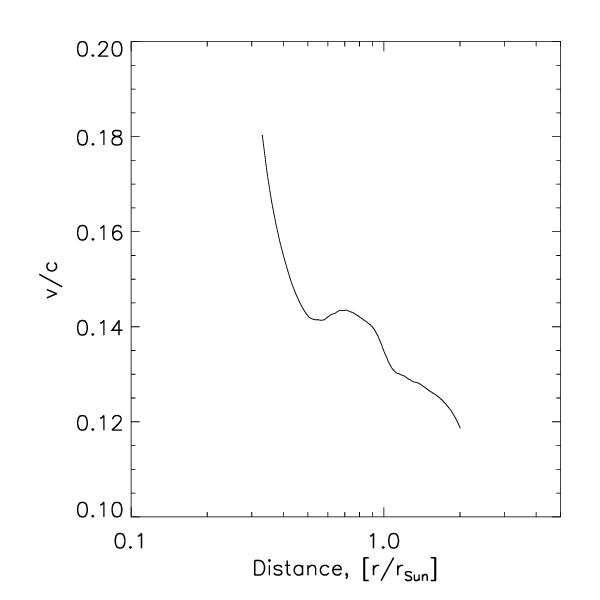}
\caption{The exciter velocity derived from the frequency drift using Equation (\ref{eqn:vEx}), as function of distance from the injection site.}
\label{fig:velyE}
\end{figure}

Observationally, empirical fits to a large number of bursts across a wide frequency range give
\begin{equation}\label{eqn:AlvHad}
df/dt \simeq -A f^{\alpha}
\end{equation} for $f$ the frequency in MHz and $df/dt$ the drift in MHz~s$^{-1}$, with $A=0.01, \alpha=1.84$ \citep{1973SoPh...29..197A, 1985srph.book.....M} for bursts between 550~MHz and 74~kHz, whereas a linear relationship of \begin{equation}df/dt = -A f +B \end{equation} with $A \sim 0.1$ and $B \sim 1$~MHz
is found by \citet{2011SoPh..269..335M} between 10 and 30 MHz similar to that found by \citet{1950AuSRA...3..541W} of \begin{equation}df/dt =- A f \end{equation} with $A=0.14 - 0.5$.

{Fitting the frequency-time curve with a single power-law across all frequencies, as shown in Figure \ref{fig:drift} gives $A=0.015$, $\alpha=1.62$. However, this fit does not work well across all frequencies. Instead, we can fit segments of the curve over small frequency ranges, giving a range of values for $A$ and $\alpha$, which vary
 with frequency within the following ranges: $A\sim 0.005-0.1$ and
 $\alpha \sim 1-2$. The resulting $df/dt$ is shown in the right panel
 of Figure \ref{fig:drift}.}

Now we can use this frequency drift to infer the source velocity, which is given by
\begin{equation}\label{eqn:vEx}
{\rm v}_{{\rm exciter}}= 2 \frac{df/dt}{f}\frac{n_e}{dn_e/dr}
\end{equation}
where $n_e$ is the background electron density at the location of the exciter. In contrast to the estimates quoted above from e.g. \citet{1950AuSRA...3..541W}, we know exactly the plasma density and its gradient from our simulations, and {the code is able to resolve the dynamic spectra in frequency and time. }
Using the known density profile we obtain source velocities as plotted in Figure \ref{fig:velyE}. Because of the fitting involved, this derived velocity may have large errors. However, there is a clear trend of deceleration with frequency. The speeds are generally in the range of inferred values \citep[e.g.][]{1987A&A...173..366D} with a value for the exciter speed of about $0.12c$ near 10~MHz. The smaller
values at lower frequency are closer to the Langmuir wave generating electrons inferred from in situ
observations near $0.02$~MHz \citep[e.g.][]{1981ApJ...251..364L}.
\citet{1992SoPh..137..307R} suggests that exciter speed should decrease
to explain the empirical results \citep{1973SoPh...29..197A, 1985srph.book.....M}.

The exciter speed inferred from the peak of electromagnetic emission differs from that derived from the peak of Langmuir wave spectral energy density, and has different frequency dependency. This can be explained by the following. The harmonic emission arises from coalescence of waves from the forwards and backwards wave populations, and so the presence of an enhanced level of backscattered Langmuir waves is an essential condition for harmonic emission. Further, there is significant spectral evolution of the Langmuir waves both due to transport effects on the generating electrons and to Langmuir wave back scattering. Thus the part of the beam which is most effective for generating emission does not correspond to the peak of Langmuir wave energy density and rather occurs where we have high levels of both forwards and backwards waves at appropriate wavenumbers.

\begin{figure}
\centering
\includegraphics[width=0.4\textwidth]{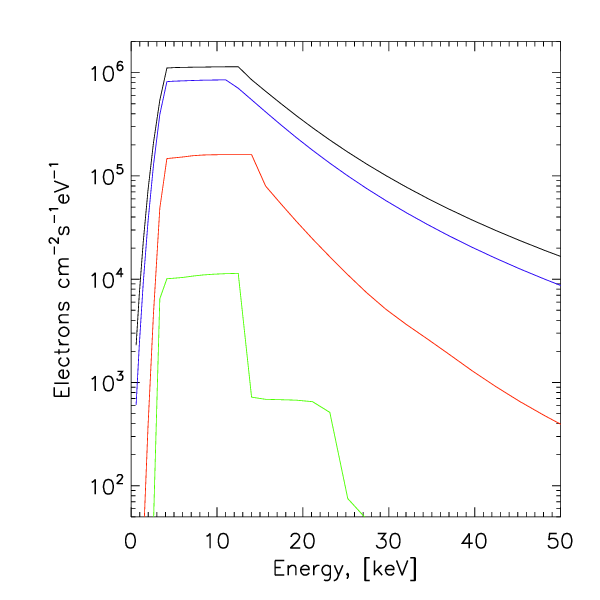}
\includegraphics[width=0.4\textwidth]{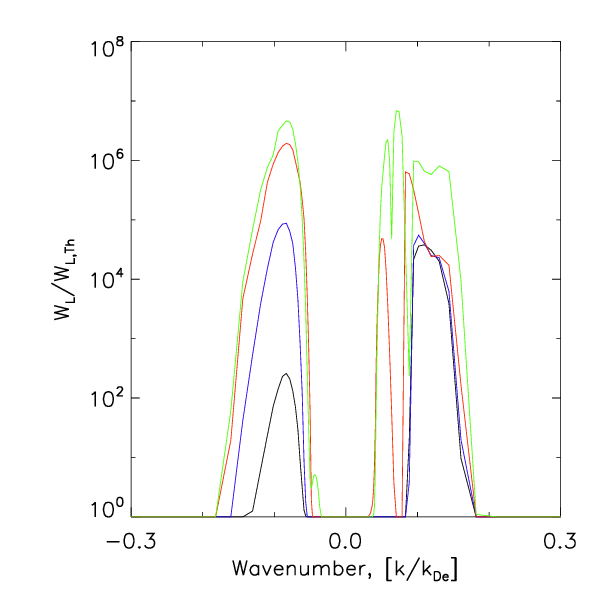}
\caption{Electron flux as a function of energy (top) and Langmuir wave spectral energy density {as a function of wavenumber normalised by the Debye wavenumber $k_{De}$} (bottom) at a local plasma frequency of 100 (black), 75 (blue), 25 (red) and 5~MHz (green) at the time of peak EM emission (as given by Figure \ref{fig:drift}).}
\label{fig:Xsec}
\end{figure}

In Figure \ref{fig:Xsec} we plot the snapshots of electron and Langmuir wave distributions at a few spatial locations. For a given local plasma frequency (100, 75, 25 and 5~MHz), we find the time at which the EM emission from this location is maximum, as shown in the left panel of Figure \ref{fig:drift}, and plot the distributions at this location and time. The onset of emission is near 200~MHz, or local plasma frequency of 100~MHz as we observed the second-harmonic component, corresponding to the black line in the Figure. At this time, the backscattered Langmuir wave level has reached several orders of magnitude over thermal, and peaks at wavenumber $k/k_{De} \sim 0.1$, corresponding to electrons at velocity $0.2c$, or energy $12$~keV. When the peak of the emission has reached 75~MHz we see a much higher level of backscattered Langmuir waves, extending to larger wavenumber, therefore resonant with slower electrons. In the next two curves, 25 and 5~MHz respectively, more Langmuir waves appear at larger, negative wavenumber, up to $k/k_{De}\sim 0.2$, corresponding to electrons at $0.1c$. This trend is the origin of the decreasing exciter velocity seen in Figure \ref{fig:velyE}.

Additionally, in the 25 and 5~MHz curves in Figure \ref{fig:Xsec} we also see multiple backscatterings producing Langmuir waves at small positive wave numbers. These then interact with faster electrons, and produce the double plateau seen in the 5~MHz electron distribution cross-section.

\section{Discussion and conclusions}

In this work we have described a model simulating electron injection in the solar corona, and the subsequent evolution and wave production which leads to radio emission. To summarise, electrons are injected with a power-law distribution in velocity at a  location corresponding
to 550~MHz and propagate along the expanding magnetic field.  The distribution of electrons evolves due to transport effects (where faster electrons outpace slower ones) to form a reverse slope in velocity space that is unstable to Langmuir wave generation. These Langmuir waves may then be re-absorbed by the beam, refracted, backscattered, and waves from the forwards and backwards populations can coalesce to produce radio emission at twice the local plasma frequency. Emission at the local plasma frequency, which can be due to scattering of Langmuir waves by ions or their interaction with ion-sound waves, is seen to be weak for the corona and beam parameters chosen.

Using the assumptions and simplifications outlined in Section 2, we have self-consistently
treated all the steps in this process to produce simulated dynamic spectra of emission,
and analysed the properties of the resulting radiation at the Earth. Our main findings are
summarised here.

An observable level of harmonic plasma emission arises over wide range of Langmuir wave levels
and different shapes of electron distribution function. A noticeable level of electromagnetic emission appears both when the flattening of the electron distribution function by quasilinear relaxation is weak near the starting frequency ($\sim200$~MHz) and when the distribution has a plateau-like shape at dekameter wavelengths ($\sim10$~MHz).

The simulations show that burst decay time is not, as has been previously suggested,
due to collisional decay of Langmuir waves. The collisional decay time is too short at high frequencies
and too long at low frequencies although for a small range around perhaps 30-100~MHz the decay time
and the collisional time are coincidentally similar. Instead, at high frequencies the duration (and decay time) depends on the duration of excitation, which is itself a function of the injection characteristics of fast electrons. Further investigation of the effects of injection time on the emission is ongoing. On the other hand at lower frequencies it appears that plasma inhomogeneity plays a crucial role,
by shifting Langmuir waves out of resonance with the fast electrons.

As discussed previously \citep[e.g.][]{2013SoPh..285..217R} we confirm that the time-injection profile of the electron beam alters the starting frequency of the radio emission, with a slower injection leading to lower starting frequency and vice versa.

The dominant factor in the instantaneous emission bandwidth is found to be the spatial extension of the emission source. At a given time, Langmuir waves are present over a large spatial extent, and therefore at a wide range of plasma frequencies. Backscattered Langmuir waves show a similar dispersion, and this is therefore mirrored in the harmonic emission. Again, this may be expected to depend on electron injection characteristics.

The frequency drift of the peak of emission can be used to infer the velocity
of the emitting source {if the background plasma density is known}. However, the exciter of type III bursts cannot be uniquely associated
with either electrons of a particular speed or with the peak of the Langmuir waves. There is a closer association of the derived exciter speed with the secondary (back-scattered) Langmuir waves,
and the velocity of the exciter is better approximated by the speed of the peak of these
secondary waves. The simulations suggest the rather slow deceleration of the exciter
as frequency decreases. Indeed, the evolution of the electron distribution
at the peak of the burst does show the distribution flattening between a few keV and around 15 keV.
The overall deceleration of the beam-plasma structure is consistent with previous results \citep[e.g.][]{2001SoPh..202..131K, 2013SoPh..285..217R} and is due to evolution of Langmuir
waves to larger wavenumbers
(resonant with slower electrons) as the beam-plasma structure moves into the interplanetary space.

To conclude, we have investigated the several large-scale characteristics of type III radio bursts using large scale 
simulations covering almost two orders in frequency. Due to the complex nature of the emission process such simulations 
are invaluable in order to determine which physical processes 
responsible for the particular features of the emission observed. Initial results have clarified several points 
regarding the burst excitation and duration, but there remains much to consider, 
especially regarding the influence of electron injection characteristics, and the time 
and frequency evolution of the emission processes.

\begin{acknowledgements}
We thank the referee for detailed comments on the manuscript. 
This work is supported by the European Research Council under the SeismoSun Research Project No. 321141 (H Ratcliffe), 
a Marie Curie International Research Staff Exchange Scheme `Radiosun'€ (PEOPLE-2011-IRSES-295272), 
an STFC consolidated grant (E.P. Kontar) and a SUPA LOFAR Advanced Fellowship (H.A.S. Reid). 
Additional thanks are due to ISSI (International Space Science Institute) for useful discussions during International Team meetings.

\end{acknowledgements}

\bibliographystyle{aa1}
\bibliography{refs1}

\end{document}